\documentclass[journal,10pt,singlespace,comsoc]{IEEEtran}
\usepackage[margin=0.642in,top=1in]{geometry}
\usepackage{graphicx, xcolor,colortbl, soul}
\usepackage{epstopdf}
\usepackage{amsfonts,amssymb,units}
\usepackage{url}
\usepackage{gensymb}
\usepackage{multicol}
\usepackage{changepage}
\usepackage[english]{babel}
\usepackage{float}
\usepackage{multirow}
\usepackage{amsmath, amssymb, bm, cite, epsfig,xfrac}
\usepackage{tabularx}
\usepackage{tabu}
\usepackage{pbox}
\usepackage{makecell}

\usepackage{booktabs}
\usepackage{ctable}
\usepackage{color}
\usepackage{booktabs}

\usepackage{graphicx}
\usepackage[caption=false]{subfig}

\def\PL{\mathrm{PL}}

\def\tm1{t\! - \! 1}
\def\tp1{t\! + \! 1}
\def\log{\mathrm{log}}

\def\log{\mathrm{log}}

\def\RF{\mathrm{RF}}
\def\S{\mathrm{S}}
\def\BB{\mathrm{BB}}

\def\Tr{\mathrm{Tr}}

\def\2D{\mathrm{2D}}
\def\3D{\mathrm{3D}}
\def\T{\mathrm{T}}
\def\R{\mathrm{R}}
\def\S{\mathrm{S}}
\def\RF{\mathrm{RF}}
\def\BB{\mathrm{BB}}

\def\SLNR{\mathrm{SLNR}}

\def\Tr{\mathrm{tr}}

\def\t{\mathrm{t}}

\definecolor{Gray}{gray}{0.85}
\definecolor{LightCyan}{rgb}{0.88,1,1}

\usepackage[absolute]{textpos}
\begin{document}
	\begin{textblock}{18}(2,0.5)
		\centering
		\noindent\Large S. Sun, T. S. Rappaport, and M. Shafi, "Hybrid beamforming for 5G millimeter-wave multi-cell networks," in \textit{Proceedings of the IEEE Conference on Computer Communications Workshops (INFOCOM WKSHPS)}, Honolulu, HI, USA, Apr. 2018.
	\end{textblock}

\title{Hybrid Beamforming for 5G Millimeter-Wave Multi-Cell Networks}

\author{\IEEEauthorblockN{Shu Sun$^*$, Theodore S. Rappaport$^*$, and Mansoor Shafi$^\dagger$}\\
	\IEEEauthorblockA{$^*$NYU WIRELESS and NYU Tandon School of Engineering, New York University, Brooklyn, NY, USA\\
		$^\dagger$Spark New Zealand, Wellington, New Zealand\\
		\{ss7152, tsr\}@nyu.edu, Mansoor.Shafi@spark.co.nz}
	
	\thanks{Sponsorship for this work was provided by the NYU WIRELESS Industrial Affiliates program and NSF research grants 1320472, 1302336, and 1555332.}
}
\maketitle

\begin{abstract}
Multi-cell wireless systems usually suffer both intra-cell and inter-cell interference, which can be mitigated via coordinated multipoint (CoMP) techniques. Previous works on multi-cell analysis for the microwave band generally consider fully digital beamforming that requires a complete radio-frequency chain behind each antenna, which is less practical for millimeter-wave (mmWave) systems where large amounts of antennas are necessary to provide sufficient beamforming gain and to enable transmission and reception of multiple data streams per user. This paper proposes four analog and digital hybrid beamforming schemes for multi-cell multi-user multi-stream mmWave communication, leveraging CoMP. Spectral efficiency performances of the proposed hybrid beamforming approaches are investigated and compared using both the 3rd Generation Partnership Project and NYUSIM channel models. Simulation results show that CoMP based on maximizing signal-to-leakage-plus-noise ratio can improve spectral efficiency as compared to the no-coordination case, and spectral efficiency gaps between different beamforming \textcolor{black}{approaches} depend on the interference level that is influenced by the cell radius and the number of users per cell.
\end{abstract}
\section{Introduction}
Millimeter-wave (mmWave) cellular systems are expected to be deployed in fifth-generation (5G) networks to achieve much greater data rates using much wider bandwidth channels. In dense networks, a major challenge that needs to be solved is inter-cell interference. Extensive research work has been done on eliminating or mitigating inter-cell interference. Power control and antenna array beamforming are two basic approaches for controlling multi-user interference~\cite{Rashid98}, but power control mainly improves the quality of weak links by equalizing the signal-to-interference-plus-noise ratio (SINR) for all users in a cell. However, antenna arrays can improve desired signal quality whilst mitigating interference by adjusting beam patterns. Antenna array beamforming is more compelling for mmWave systems as compared to power control since antenna arrays are expected to be used at both communication link ends to provide array gain to compensate for the higher free space path loss \textcolor{black}{in the first meter of propagation}. To reduce interference using antenna arrays, one promising solution is letting base stations (BSs) or transmission points (TPs) in different cells cooperate in transmission and/or reception using antenna arrays. 

The 3rd Generation Partnership Project (3GPP) completed a study on coordinated multipoint (CoMP) techniques for the fourth-generation (4G) Long Term Evolution (LTE)-Advanced system in 2013~\cite{3GPP_36.819}. Different CoMP strategies in~\cite{3GPP_36.819} entail different levels of complexity and requirements with respect to channel state information (CSI) feedback and CSI sharing, which are detailed below in increasing order of complexity and requirements. 

\subsubsection{Coordinated Scheduling/Beamforming}
Data for a \textcolor{black}{mobile} user equipment (UE) is only available at and transmitted from one \textcolor{black}{TP} in the CoMP cooperating set (downlink data transmission is done from that \textcolor{black}{specific TP}) for a time-frequency resource, but user scheduling/beamforming decisions are made with coordination among multiple TPs. 

\subsubsection{Dynamic Point Selection (DPS)/Muting}
Data is available simultaneously at multiple TPs but is transmitted from only one \textcolor{black}{TP} in a time-frequency resource using its own beamforming approach, and the transmitting/muting \textcolor{black}{TP} may change from one subframe \textcolor{black}{(time or frequency resource)} to another. 

\subsubsection{Joint transmission}
Data for a UE is available at multiple TPs and is simultaneously transmitted from multiple TPs to a single UE or multiple UEs in \textcolor{black}{the same} time-frequency resource. 

BS coordination for interference suppression has been extensively explored in the literature in the past decade, such as the works in~\cite{Lee12_Feb,Schwarz14,Sadek07,Yang17_TWC}, yet those works focused on fully digital beamforming with one radio-frequency (RF) chain behind each antenna, which is not likely to be suitable for mmWave systems with large amounts (e.g., hundreds) of antennas at BSs due to hardware complexity, power consumption, and cost. BS cooperation in mmWave multi-cell networks was investigated in~\cite{Maamari16,Muh17_ICC,Zhu17_JSAC}, but the mobile receiver was equipped with merely a single \textcolor{black}{omnidirectional} antenna hence leading to only single-stream communication in those works. In 5G mmWave systems, however, antenna arrays will also be employed at the mobile receiver to provide array gain and beamforming and/or spatial multiplexing capability.

In this paper, we investigate multi-cell multi-user \textit{multi-stream} analog and digital hybrid beamforming (HBF) strategies for mmWave multiple-input multiple-output (MIMO) systems using four schemes: three that use coordinated scheduling/beamforming, and one that does not use any TP coordination (as a baseline), which has not been studied before to our best knowledge. In this work, we focus on the forward link from the TP to the UE, and assume equal power allocations are used for each stream (i.e. no \textcolor{black}{power control or water filling per stream}). A multi-cell framework is formulated based upon today's conventional three-sector BS antenna configuration, where each 120$^\circ$ sector (i.e., cell, as defined in 3GPP parlance~\cite{3GPP_36.819}) uses a uniform rectangular array (URA) with 256 antenna elements (eight rows by 16 columns by two polarizations) for each TP, similar to what is envisioned for 5G MIMO systems\textcolor{black}{~\cite{M2412,Vook16}}. The spacing between adjacent co-polarized elements is $\lambda/2$ \textcolor{black}{in azimuth and $\lambda$ in elevation where} $\lambda$ denotes the carrier wavelength (e.g., 10.7 mm at 28 GHz and 4.1 mm at 73 GHz), and the radiation pattern of each antenna element is given in Table~\ref{tbl:SimSet}, which provides a 3 dB beamwidth resolution of about 8$^\circ$ in the broadside direction of the URA \textcolor{black}{at each TP}. Note that the number of RF chains used to feed the URA dictates the \textcolor{black}{maximum} number of independent RF streams that may be transmitted \textcolor{black}{but shared over} all users in a cell. A number of (3 or 12 in this work) UEs, each with an eight-element URA and four RF chains (for up to four streams per user), are randomly dropped in each cell over distances \textcolor{black}{ranging between} 10 m \textcolor{black}{and} the cell radius (e.g., 50 m or 200 m), and 100 MHz channel bandwidths are used assuming orthogonal frequency-division multiplexing (OFDM)-like (single channel per tone) modulation with small channel bandwidths for flat fading. \textcolor{black}{5G systems will have large bandwidths (e.g., 1 GHz), but this bandwidth is likely to be aggregated over RF channels which are 100 MHz wide and which use many OFDM sub-carriers that are each narrowband (flat-fading) in nature~\cite{Vook16,Rap15}.} URAs are considered because they are able to form beams in both azimuth and elevation dimensions, as will be required in 5G mmWave systems\textcolor{black}{~\cite{M2412}}. It is assumed that the TPs in different cells (i.e., 120$^\circ$ sectors) have full CSI and can exchange the CSI among each other, such that TPs can take actions to mitigate inter-cell interference, which corresponds to coordinated scheduling/beamforming per the definition by 3GPP~\cite{3GPP_36.819}. The main contributions and observations of this paper are as follows:
\begin{itemize}
	\item Four multi-cell HBF approaches are proposed and compared in terms of spectral efficiency under various conditions (e.g., different cell radii, \textcolor{black}{numbers of users}, and \textcolor{black}{numbers of streams} per user), using both the 3GPP TR 38.901 Release 14 channel model~\cite{3GPP_38.901} and the NYUSIM channel model~\cite{Sun17_NYUSIM}. 
	\item Inter-cell TP coordination based on a strategy that maximizes signal-to-leakage-plus-noise ratio (SLNR) for each user in every cell is shown to improve spectral efficiency by as much as 67\% when compared to the no-coordination case, where \textit{leakage} refers to the amount of interference caused by the signal intended for a desired user but received by the remaining users \textcolor{black}{in all cells considered}, in contrast to \textit{interference} that is generated from undesired TPs and received by the desired user~\cite{Sadek07,Feng11,Bouk17}. Furthermore, we show that the SLNR-based approach can virtually eliminate interference for each user when each cell is lightly loaded (e.g., three users per cell).
	\item For the same cell radius and the same forward-link transmit power for each user \textcolor{black}{without power control}, an increase in the \textcolor{black}{number of users} per cell results in lower per-user spectral efficiency due to the increased inter-user interference.
	\item For the same \textcolor{black}{number of users} per cell and the same forward-link transmit power for each user \textcolor{black}{without power control}, a smaller cell radius leads to higher per-user spectral efficiency in most cases, primarily due to the enhanced received signal power (i.e., lower path loss) from smaller transmitter-receiver (T-R) separation distances.
\end{itemize}
\begin{table}\footnotesize
	\renewcommand{\arraystretch}{1}
	\centering
	\caption{Simulation settings using 3GPP~\cite{3GPP_38.901} and NYUSIM~\cite{Sun17_NYUSIM} models.}\label{tbl:SimSet}
	\newcommand{\tabincell}[2]{\begin{tabular}{@{}#1@{}}#2\end{tabular}}
	\begin{tabular}{|c|c|}
		\hline
		\tabincell{c}{\textbf{Parameter}}&\tabincell{c}{\textbf{Setting}}\\
		\Xhline{1.5pt}
		\tabincell{c}{\textbf{Carrier Frequency}}&28 GHz\\
		\hline
		\tabincell{c}{\textbf{Bandwidth}}&100 MHz\\
		\hline
		\tabincell{c}{\textbf{Transmit Power}}&\tabincell{c}{35.2 dBm per UE}\\
		\hline
		\textbf{\tabincell{c}{95\% Cell-Edge SNR}}&\tabincell{c}{5 dB}\\
		\hline
		\textbf{\tabincell{c}{BS Antennas}}&\tabincell{c}{three panels for the three \textcolor{black}{TP} sectors, \\where each panel is a uniform \\rectangular array consisting of \textcolor{black}{256} \\cross-polarized elements in the \\x-z plane}\\
		\hline
		\textbf{\tabincell{c}{BS Antenna Spacing}}&\tabincell{c}{half wavelength in azimuth,\\one wavelength in elevation}\\
		\hline	
		\textbf{\tabincell{c}{BS Antenna Element Gain}}&\tabincell{c}{8 dBi~\cite{3GPP_38.901}}\\
		\hline
		\textbf{\tabincell{c}{BS Antenna Element \\Pattern}}&\tabincell{c}{Model 2, Page 18 in 3GPP \\TR 36.873 Release 12~\cite{3GPP_36.873}}\\
		\hline
		\textbf{\tabincell{c}{UE Antennas}}&\tabincell{c}{uniform rectangular array \\consisting of \textcolor{black}{8} cross-polarized \\elements in the x-z plane}\\
		\hline
		\textbf{\tabincell{c}{UE Antenna Spacing}}&\tabincell{c}{half wavelength in azimuth,\\one wavelength in elevation}\\
		\hline	
		\textbf{\tabincell{c}{UE Antenna Element Gain}}&\tabincell{c}{0 dBi}\\
		\hline
		\textbf{\tabincell{c}{UE Antenna Element Pattern}}&\tabincell{c}{omnidirectional}\\
		\hline	
		\textbf{\tabincell{c}{Receiver Noise Figure}}&\tabincell{c}{10 dB}\\
		\hline	
	\end{tabular}
\vspace{-4 mm}
\end{table}
\vspace{-2 mm}
\section{Multi-Cell System Layout and Hybrid Beamforming Framework}\label{sec:MCSL}
\begin{figure}
	\centering
	\includegraphics[width=2.3in]{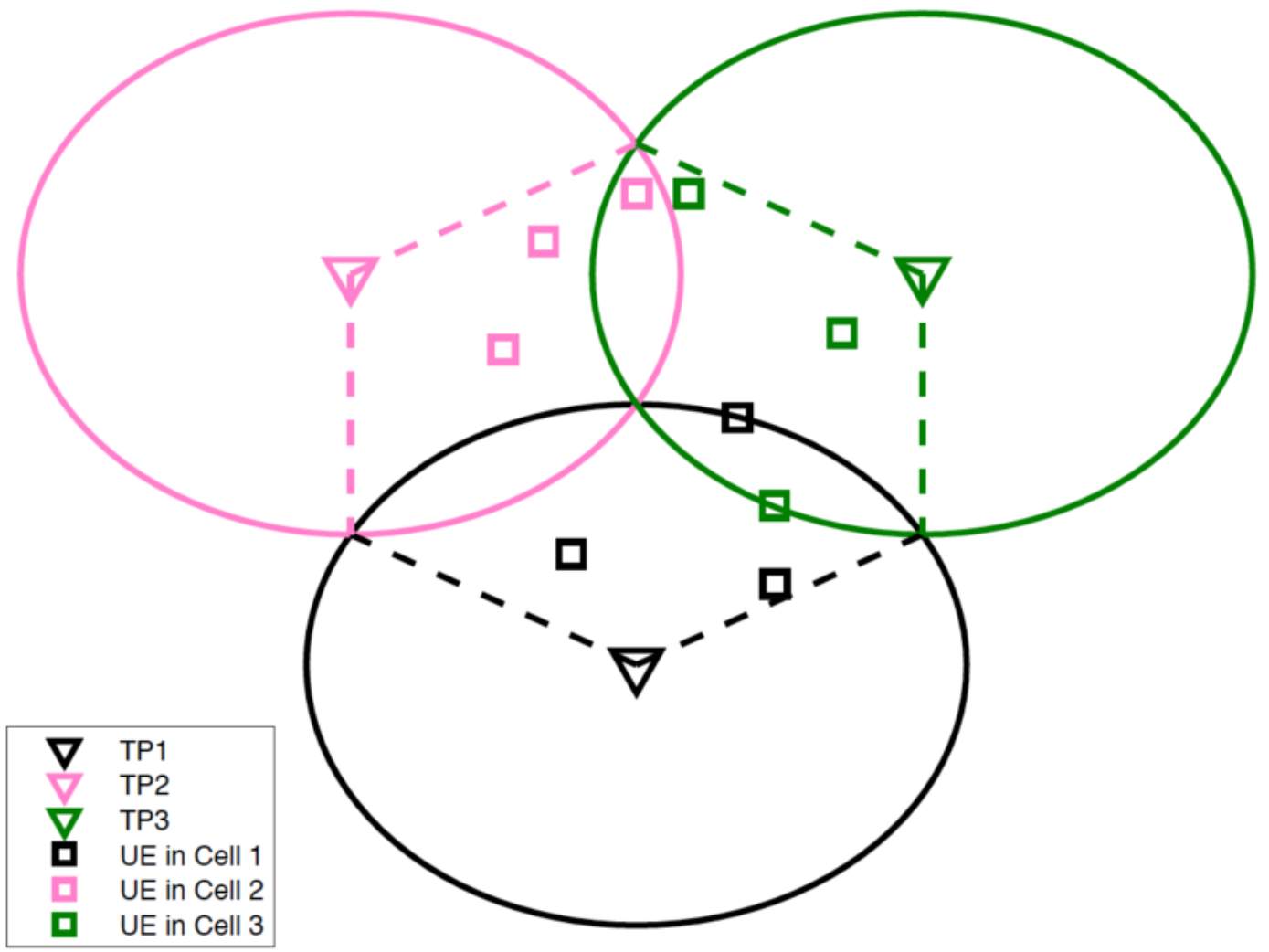}
	\vspace{-3 mm}
	\caption{An example of the three-cell layout where there is one TP and three UEs per cell generated using MATLAB, where each cell is a sector with an azimuth span of 120$^\circ$ served by one TP, and UEs in each cell are dropped randomly and uniformly with T-R separation distances ranging from 10 m to the cell radius (e.g., 50 m or 200 m).}
	\label{fig:Cells}
	\vspace{-4 mm}
\end{figure}

\vspace{-0.5 mm}
We consider an mmWave system with three adjacent cells (i.e., sectors), each having one TP and multiple (e.g., 3 or 12) UEs\textcolor{black}{, referred to as a coordination cluster}. Only three \textcolor{black}{adjacent} cells are studied herein since inter-cell interference among these three cells \textcolor{black}{will} dominate the interference due to the geographical proximity and use of mmWave frequencies. \textcolor{black}{Further, the antenna element is modeled with a sectoral antenna pattern~\cite{3GPP_36.873}, and the array has the required array pattern (e.g., about 8$^\circ$ 3 dB beamwidth), so that users out of the sectoral range do not see the benefit of the array. Therefore, the three-cell system} is representative of homogeneous multi-cell networks with both intra- and inter-cell interference. \textcolor{black}{The} four proposed HBF approaches are applicable to general cases with more cells. Fig.~\ref{fig:Cells} depicts an example of the three-cell layout with three users per cell. \textcolor{black}{Interference from neighboring coordination clusters is ignored in this work. Inclusion of the interference from neighboring clusters will lower the SINR for all beamforming approaches.}

\section{Multi-Cell Multi-User Multi-Stream Hybrid Beamforming}
\begin{figure}
	\centering
	\includegraphics[width=2.6in]{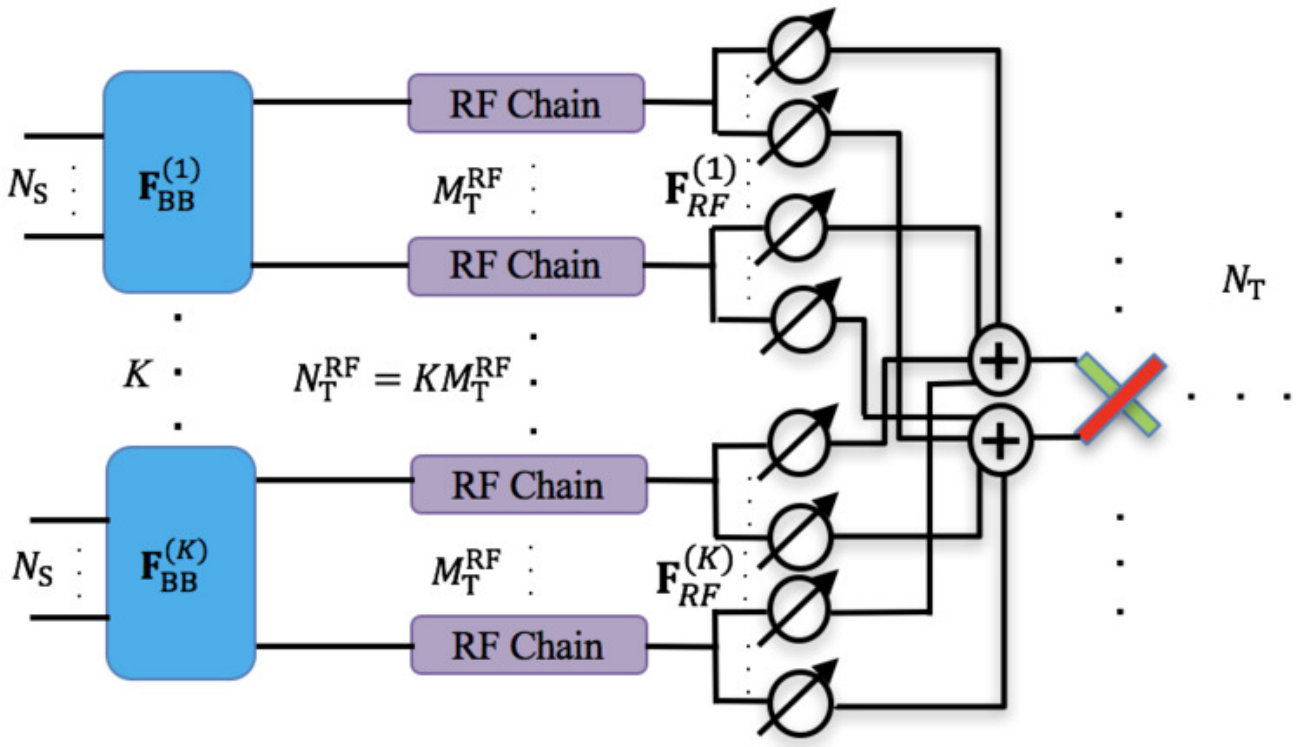}
	\vspace{-3 mm}
	\caption{Multi-cell HBF architecture at the TP in each cell (there are three TPs in one BS, and one TP serves one cell). \textcolor{black}{$N_\S$ denotes the number of data streams per user in each cell, $K$ is the number of users in each cell, $N_\T^{\RF}$ represents the total number of RF chains at each TP, $M_\T^{\RF}$ is the number of RF chains connected to the baseband precoder for one user, and $N_\T$ denotes the number of TP antenna elements in each cell. In this multi-cell multi-stream work, $N_\S$ varies from 1 to 4, $K$ is either 3 or 12, $M_\T^{\RF}=4$ which equals the number of RF chains at each UE, $N_\T^{\RF}=KM_\T^{\RF}$ which is either 12 or 48, and $N_\T=256$.}}
	\label{fig:MultiCellHBFStructure2}
	\vspace{-3 mm}	
\end{figure}
\begin{figure}
	\centering
	\includegraphics[width=2.3in]{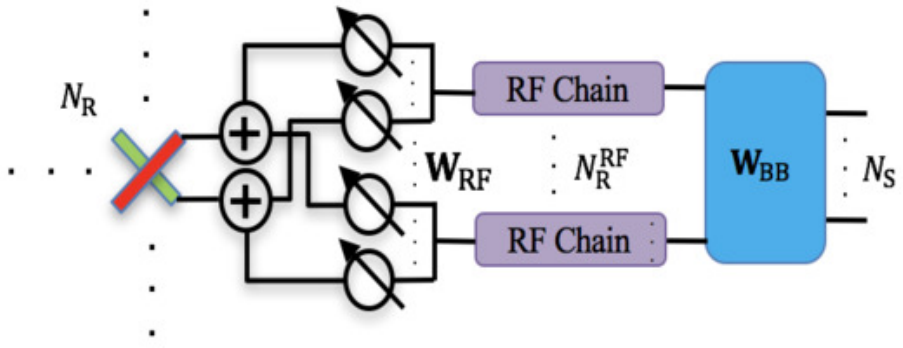}
	\vspace{-3 mm}
	\caption{\textcolor{black}{Multi-cell HBF architecture at each UE. $N_\S$ denotes the number of data streams per UE, $N_\R^{\RF}$ represents the number of RF chains at each UE, and $N_\R$ denotes the number of UE antenna elements. In this multi-cell multi-stream work, $N_\S$ varies from 1 to 4, $N_\R^{\RF}=4$, and $N_\R=8$.}}
	\label{fig:MultiCellHBFStructure_UE}	
	\vspace{-3 mm}
\end{figure}

Consider Fig.~\ref{fig:Cells} where each cell has one TP equipped with an $N_\T=256$ element URA (eight rows by 16 columns by two polarizations), and multiple users each with an $N_\R=8$ element URA (two rows by two columns by two polarizations). The HBF architecture in Fig.~\ref{fig:MultiCellHBFStructure2} is used at each TP, where the RF chains are divided into $K$ subsets with $M_\T^{\RF}$ (fixed at four in this work) RF chains in each subset, such that the total number of RF chains is $N_\T^{\RF}=KM_\T^{\RF}$ where $K$ is 3 or 12 here. Additionally, at each TP, there are $K$ baseband digital precoders each connected to a subset dedicated to a user in the home cell. \textcolor{black}{The URA architecture at each UE is illustrated in Fig.~\ref{fig:MultiCellHBFStructure_UE}, where there are $N_\R$ antennas and $N_\R^{\RF}$ RF chains at each UE, and all the RF chains are connected to all the antennas.} The approaches in this work assume all UEs use all four RF chains, even if the stream number is less than four. For TP $i$ and user $k$ in cell $l$, the $N_\R\times N_\T$ downlink channel is denoted as \textbf{H}$_{k,l,i}$, the $N_\T\times M_\T^{\RF}$ RF precoding matrix is $\textbf{F}_{\RF_{k,l}}$, and the $M_\T^{\RF}\times N_\S$ baseband precoding matrix is \textbf{F}$_{\BB_{k,l}}$. The $N_\R\times N_\R^{\RF}$ RF combining matrix and the $N_\R^{\RF}\times N_\S$ baseband combining matrix are \textbf{W}$_{\RF_{k,l}}$ and \textbf{W}$_{\BB_{k,l}}$, respectively. The received signal at user $k$ in cell $l$ can be formulated as:
\begin{equation}\label{eq:yMultiStream}\footnotesize
\begin{split}
\textbf{y}_{k,l}=&\underbrace{\sqrt{\frac{P_\t}{\eta_{k,l}\PL_{k,l,l}}}\textbf{W}_{\BB_{k,l}}^H\textbf{W}_{\RF_{k,l}}^H\textbf{H}_{k,l,l}\textbf{F}_{\RF_{k,l}}\textbf{F}_{\BB_{k,l}}\textbf{s}_{k,l}}_{\textrm{Desired~Signal}}\\
&+\underbrace{\sum_{\substack{(m,i)\\\neq(k,l)}}\sqrt{\frac{P_\t}{\eta_{m,i}\PL_{k,l,i}}}\textbf{W}_{\BB_{k,l}}^H\textbf{W}_{\RF_{k,l}}^H\textbf{H}_{k,l,i}\textbf{F}_{\RF_{m,i}}\textbf{F}_{\BB_{m,i}}\textbf{s}_{m,i}}_{\textrm{Interference}}\\
&+\underbrace{\textbf{W}_{\BB_{k,l}}^H\textbf{W}_{\RF_{k,l}}^H\textbf{n}_{k,l}}_{\textrm{Noise}}
\end{split}
\end{equation}

\noindent where $P_\t$ represents the transmit power for each user in Watts, and is assumed to be constant regardless of the \textcolor{black}{number of users} per cell and the cell radius. $\PL_{k,l,i}$ denotes the large-scale distance-dependent path loss in Watts, including shadow fading, from TP $i$ to user $k$ in cell $l$, $\eta_{k,l}=||\textbf{F}_{\RF_{k,l}}\textbf{F}_{\BB_{k,l}}||_F^2$ is a scaling factor to satisfy the per-user transmit power constraint $||\sqrt{P_\t}\textbf{F}_{\RF_{k,l}}\textbf{F}_{\BB_{k,l}}/\sqrt{\eta_{k,l}}||_F^2=P_\t$, where $F$ denotes the Frobenius norm. \textbf{s}$_{k,l}$ represents the desired transmitted signal for user $k$ in cell $l$ with $\mathbb{E}[\textbf{s}_{k,l}\textbf{s}_{k,l}^H]=\textbf{I}_{N_\S}$, and $\textbf{n}_{k,l}\sim\mathcal{CN}(\bold{0},N_0\textbf{I}_{N_\R})$ is circularly symmetric complex Gaussian noise with variance $N_0$. The $N_\R^{\RF}\times M_\T^{\RF}$ effective channel $\check{\textbf{H}}_{k,l,m,i}$ after RF precoding and RF combining is:
\begin{equation}\label{eq:Heff}
\begin{split}
\check{\textbf{H}}_{k,l,m,i}=\textbf{W}_{\RF_{k,l}}^H\textbf{H}_{k,l,i}\textbf{F}_{\RF_{m,i}}
\end{split}
\end{equation}

\noindent The spectral efficiency of user $k$ in cell $l$ is calculated as in~\eqref{eq:R}\textcolor{black}{~\cite{Ayach14}}, where the interference term $\textbf{D}$ in~\eqref{eq:R} is given by:
\begin{figure*}
	\begin{equation}\label{eq:R}
	\begin{split}
	R_{k,l}=&\log_2\Bigg|\textbf{I}_{N_\S}+\frac{P_\t}{\eta_{k,l}\PL_{k,l,l}}\big(\textbf{W}_{\BB_{k,l}}^H\textbf{W}_{\RF_{k,l}}^H(N_0\textbf{I}_{N_\R}+\textbf{D})\textbf{W}_{\RF_{k,l}}\textbf{W}_{\BB_{k,l}}\big)^{-1}\textbf{W}_{\BB_{k,l}}^H\breve{\textbf{H}}_{k,l,k,l}\textbf{F}_{\BB_{k,l}}\textbf{F}_{\BB_{k,l}}^H\breve{\textbf{H}}_{k,l,k,l}^H\textbf{W}_{\BB_{k,l}}\Bigg|\\
	\end{split}
	\end{equation}
	\hrulefill
\end{figure*}
\begin{equation}\label{eq:D1}
\textbf{D}=\sum_{\substack{(m,i)\\\neq(k,l)}}\frac{P_\t}{\eta_{m,i}\PL_{k,l,i}}\textbf{H}_{k,l,i}\textbf{F}_{\RF_{m,i}}\textbf{F}_{\BB_{m,i}}\textbf{F}_{\BB_{m,i}}^H\textbf{F}_{\RF_{m,i}}^H\textbf{H}_{k,l,i}^H
\end{equation}

\noindent \textcolor{black}{Note that the spectral efficiency in~\eqref{eq:R} is formulated based on Shannon theory assuming ideal encoding and decoding functions and serves as an upper bound of the achievable rate~\cite{Shannon48}. Non-ideal/more practical encoding and decoding may be used in reality which results in lower spectral efficiency compared to~\eqref{eq:R}. Additionally, for all the multi-cell HBF approaches henceforth, it is assumed that no power control is performed.}
\vspace{-3 mm}
\subsection{Baseline Case --- No Coordination Among Cells}
Let us first consider the interference-ignorant baseline case where there is no TP coordination among cells. Assuming only local CSI is available at each TP, a reasonable precoding scheme is eigenmode transmission\textcolor{black}{~\cite{Telatar99}. User $k$ in cell $l$ will be treated as the desired user in all the subsequent multi-cell HBF design.} Let us define the effective channel matrix $\check{\textbf{H}}_{k,l,k,l}\in\mathbb{C}^{N_\R^{\RF}\times M_\T^{\RF}}$ for user $k$ in cell $l$ as $\frac{1}{\sqrt{\PL_{k,l,l}}}\textbf{W}_{\RF_{k,l}}^H\textbf{H}_{k,l,l}\textbf{F}_{\RF_{k,l}}$, where the RF precoding and RF combining matrices \textbf{F}$_{\RF_{k,l}}$ and~\textbf{W}$_{\RF_{k,l}}$ are designed such that $||\textbf{W}_{\RF_{k,l}}^H\textbf{H}_{k,l,l}\textbf{F}_{\RF_{k,l}}||_F^2$ is maximized \textcolor{black}{to enhance signal-to-noise ratio (SNR)}. The RF beamforming approach in Eqs. (12)-(14) proposed in~\cite{Song16} is applied to obtain \textbf{F}$_{\RF_{k,l}}$ and \textbf{W}$_{\RF_{k,l}}$, in which the codebooks for \textbf{F}$_{\RF_{k,l}}$ and \textbf{W}$_{\RF_{k,l}}$ consist of the TP and UE antenna array response vectors corresponding to the angles-of-departure (AoDs) and angles-of-arrival (AoAs) associated with the desired user, respectively~\cite{Ayach14}. The baseband precoding matrix $\textbf{F}_{\BB_{k,l}}$ is composed of the dominant $N_\S$ right singular vectors obtained from the singular value decomposition (SVD) of $\check{\textbf{H}}_{k,l,k,l}$, and the baseband combining matrix $\textbf{W}_{\BB_{k,l}}$ is constituted by the dominant $N_\S$ left singular vectors obtained from the SVD of $\check{\textbf{H}}_{k,l,k,l}\textbf{F}_{\BB_{k,l}}$.
\vspace{-3 mm}
\subsection{Leakage-Suppressing and Signal-Maximizing Precoding}
\textcolor{black}{A coordinated scheduling/beamforming CoMP scheme named leakage-suppressing and signal-maximizing precoding (LSP) is proposed herein, where} the RF precoder is aimed at mitigating the dominant leakage to all the other users while enhancing the strength of the desired signal. \textcolor{black}{The precoding matrix at TP $l$ for user $k$ in cell $l$ is designed as follows.} First, the cascaded \textcolor{black}{leakage} channel matrix consisting of all the channel matrices except the one for user $k$ in cell $l$ is obtained through CSI exchange among TPs as:
\begin{equation}\label{eq:tilH1}
\begin{split}
	\tilde{\textbf{H}}_{k,l}=&\Bigg[\frac{1}{\sqrt{\PL_{1,1,l}}}\textbf{H}_{1,1,l}^T, ..., \frac{1}{\sqrt{\PL_{k-1,l,l}}}\textbf{H}_{k-1,l,l}^T,\\
	& \frac{1}{\sqrt{\PL_{k+1,l,l}}}\textbf{H}_{k+1,l,l}^T,..., \frac{1}{\sqrt{\PL_{K,L,l}}}\textbf{H}_{K,L,l}^T\Bigg]^T
	\end{split}
\end{equation}

\noindent The columns of RF beamforming matrices at each TP and UE are selected from pre-defined beamforming codebooks that consist of antenna array response vectors $\textbf{a}_{\T}$ and $\textbf{a}_{\R}$ at the TP and UE, respectively. The matrix $\textbf{A}_\T$ and $\textbf{A}_\R$ are composed of $\textbf{a}_{\T}$'s and $\textbf{a}_{\R}$'s corresponding to the AoDs and AoAs associated with the desired user, respectively~\cite{Ayach14}. The first column in the RF precoding matrix $\textbf{F}_{\RF_{k,l}}$ is chosen from $\textbf{A}_{\T}$ such that $||\tilde{\textbf{H}}_{k,l}\textbf{F}_{\RF_{k,l}}(:,1)||^2_F$ is minimized, whose physical meaning is using the first RF precoding vector \textcolor{black}{at TP $l$ to minimize} the leakage to all the other users \textcolor{black}{in all the cells considered}. The remaining $M_\T^\RF-1$ columns in $\textbf{F}_{\RF_{k,l}}$ are selected from $\textbf{A}_{\T}$ to maximize $||\textbf{H}_{k,l,l}\textbf{F}_{\RF_{k,l}}(:,2:M_\T^\RF)||^2_F$, \textcolor{black}{i,e,} utilizing the remaining $M_\T^\RF-1$ RF precoding vectors to maximize the desired signal power to user $k$ in cell $l$. Then the baseband precoding matrix $\textbf{F}_{\BB_{k,l}}$ is designed by taking the SVD of $\textbf{H}_{k,l,l}\textbf{F}_{\RF_{k,l}}$ and setting $\textbf{F}_{\BB_{k,l}}$ as $\textbf{V}(:,1:N_\S)$ where $\textbf{V}(:,1:N_\S)$ represents the dominant $N_\S$ right singular vectors of $\textbf{H}_{k,l,l}\textbf{F}_{\RF_{k,l}}$. 

For the design of the hybrid combining matrix at user $k$ in cell $l$, \textcolor{black}{first,} the optimum fully digital combining matrix is obtained by taking the SVD of $\textbf{H}_{k,l,l}\textbf{F}_{\RF_{k,l}}\textbf{F}_{\BB_{k,l}}$, and setting the columns of the combining matrix to be the dominant $N_\S$ left singular vectors. Then the RF and baseband combining matrices are designed similarly to Algorithm 1 on Page 1505 of~\cite{Ayach14} \textcolor{black}{using} the optimum fully digital combining matrix. \textcolor{black}{As extensions of LSP, if sufficient channel diversity exists, more than one precoding vector could be used for suppressing leakage when designing the precoding matrix at each TP.}
\vspace{-4 mm}
\subsection{SLNR-Based Precoding}
\textcolor{black}{The third multi-cell HBF strategy is an SLNR-based scheme incorporating coordinated scheduling/beamforming in CoMP.} Directly maximizing the SINR involves a challenging optimization problem with coupled variables, thus the SLNR is utilized as an alternative optimization criterion. In the SLNR-based CoMP scheme, the effective channel matrix $\check{\textbf{H}}_{m,i,k,l}\in\mathbb{C}^{N_\R^{\RF}\times M_\T^{\RF}}$ is defined as $\frac{1}{\sqrt{\PL_{m,i,l}}}\textbf{W}_{\RF_{m,i}}^H\textbf{H}_{m,i,l}\textbf{F}_{\RF_{k,l}}$, and the $(KL-1)N_\R^{\RF}\times M_\T^{\RF}$ leakage matrix \textcolor{black}{for TP $l$ communicating with user $k$ in cell $l$} is given by:
\begin{equation}\label{eq:tilH2}
	\tilde{\textbf{H}}_{k,l}=\Big[\check{\textbf{H}}_{1,1,k,l}^T, ..., \check{\textbf{H}}_{k-1,l,k,l}^T, \check{\textbf{H}}_{k+1,l,k,l}^T,..., \check{\textbf{H}}_{K,L,k,l}^T\Big]^T
\end{equation}

\noindent The RF precoding and RF combining matrices \textbf{F}$_{\RF_{k,l}}$ and~\textbf{W}$_{\RF_{k,l}}$ are designed such that $||\textbf{W}_{\RF_{k,l}}^H\textbf{H}_{k,l,l}\textbf{F}_{\RF_{k,l}}||_F^2$ is maximized, where $\textbf{F}_{\RF_{k,l}}$ and $\textbf{W}_{\RF_{k,l}}$ are obtained in the same manner as in the baseline case. The baseband precoding matrix $\textbf{F}_{\BB_{k,l}}$ is designed to maximize the SLNR as follows\textcolor{black}{~\cite{Sadek07}}. The expected received signal power prior to the baseband combining process is $\mathbb{E}\Big[\frac{P_\t}{\eta_{k,l}}\textbf{s}_{k,l}^H\textbf{F}_{\BB_{k,l}}^H\breve{\textbf{H}}_{k,l,k,l}^H\breve{\textbf{H}}_{k,l,k,l}\textbf{F}_{\BB_{k,l}}\textbf{s}_{k,l}\Big]$, the expected leakage power is $\mathbb{E}\Bigg[\sum\limits_{(m,i)\neq(k,l)}\frac{P_\t}{\eta_{k,l}}\textbf{s}_{k,l}^H\textbf{F}_{\BB_{k,l}}^H\breve{\textbf{H}}_{m,i,k,l}^H\breve{\textbf{H}}_{m,i,k,l}\textbf{F}_{\BB_{k,l}}\textbf{s}_{k,l}\Bigg]$, and the expected noise power is $\mathbb{E}\big[\textbf{n}_{k,l}^H\textbf{W}_{\RF_{k,l}}\textbf{W}_{\RF_{k,l}}^H\textbf{n}_{k,l}\big]$. The SLNR is hence formulated as in~\eqref{eq:SLNR2}~\cite{Sadek07},
\begin{figure*}
\begin{equation}\label{eq:SLNR2}
	\begin{split}
		\SLNR &\approx\frac{\mathbb{E}\Big[\frac{P_\t}{\eta_{k,l}}\textbf{s}_{k,l}^H\textbf{F}_{\BB_{k,l}}^H\breve{\textbf{H}}_{k,l,k,l}^H\breve{\textbf{H}}_{k,l,k,l}\textbf{F}_{\BB_{k,l}}\textbf{s}_{k,l}\Big]}{\mathbb{E}\Bigg[\sum\limits_{(m,i)\neq(k,l)}\frac{P_\t}{\eta_{k,l}}\textbf{s}_{k,l}^H\textbf{F}_{\BB_{k,l}}^H\breve{\textbf{H}}_{m,i,k,l}^H\breve{\textbf{H}}_{m,i,k,l}\textbf{F}_{\BB_{k,l}}\textbf{s}_{k,l}\Bigg]+\mathbb{E}\big[\textbf{n}_{k,l}^H\textbf{W}_{\RF_{k,l}}\textbf{W}_{\RF_{k,l}}^H\textbf{n}_{k,l}\big]}\\ &=\frac{\Tr\Big(\frac{P_\t}{\eta_{k,l}}\textbf{F}_{\BB_{k,l}}^H\breve{\textbf{H}}_{k,l,k,l}^H\breve{\textbf{H}}_{k,l,k,l}\textbf{F}_{\BB_{k,l}}\Big)}{\Tr\Bigg(\sum\limits_{\substack{(m,i)\neq(k,l)}}\frac{P_\t}{\eta_{k,l}}\textbf{F}_{\BB_{k,l}}^H\breve{\textbf{H}}_{m,i,k,l}^H\breve{\textbf{H}}_{m,i,k,l}\textbf{F}_{\BB_{k,l}}\Bigg)+N_0\Tr\Big(\textbf{W}_{\RF_{k,l}}\textbf{W}_{\RF_{k,l}}^H\Big)}\\
		&=\frac{\Tr\Big(\textbf{F}_{\BB_{k,l}}^H\breve{\textbf{H}}_{k,l,k,l}^H\breve{\textbf{H}}_{k,l,k,l}\textbf{F}_{\BB_{k,l}}\Big)}{\Tr\Big(\textbf{F}_{\BB_{k,l}}^H\tilde{\textbf{H}}_{k,l}^H\tilde{\textbf{H}}_{k,l}\textbf{F}_{\BB_{k,l}}\Big)+\frac{\eta_{k,l}}{P_\t}N_0\Tr\Big(\textbf{W}_{\RF_{k,l}}\textbf{W}_{\RF_{k,l}}^H\Big)}\textcolor{black}{=\frac{\Tr\Big(\textbf{F}_{\BB_{k,l}}^H\breve{\textbf{H}}_{k,l,k,l}^H\breve{\textbf{H}}_{k,l,k,l}\textbf{F}_{\BB_{k,l}}\Big)}{\Tr\bigg(\textbf{F}_{\BB_{k,l}}^H\Big(\tilde{\textbf{H}}_{k,l}^H\tilde{\textbf{H}}_{k,l}+\gamma\textbf{I}_{M_\T^{\RF}}\Big)\textbf{F}_{\BB_{k,l}}\bigg)}}
	\end{split}
\end{equation}
\hrulefill
\end{figure*}
where $\tilde{\textbf{H}}_{k,l}$ is given by~\eqref{eq:tilH2}, and the second equality in~\eqref{eq:SLNR2} holds since $\mathbb{E}[\textbf{s}_{k,l}\textbf{s}_{k,l}^H]=\textbf{I}_{N_\S}$ and $\mathbb{E}[\textbf{n}_{k,l}\textbf{n}_{k,l}^H]=N_0\textbf{I}_{N_\R}$. \textcolor{black}{And $\gamma$ satisfies:}
\begin{equation}
\Tr(\gamma\textbf{F}_{\BB_{k,l}}^H\textbf{F}_{\BB_{k,l}})=\frac{\eta_{k,l}}{P_\textrm{t}}N_0\Tr(\textbf{W}_{\RF_{k,l}}\textbf{W}_{\RF_{k,l}}^H)
\end{equation}

\noindent The optimal $\textbf{F}_{\BB_{k,l}}$ that maximizes the SLNR in~\eqref{eq:SLNR2} can be derived similarly to the precoding matrix in~\cite{Sadek07} and is composed of the leading $N_\S$ columns of $\textbf{T}_{k,l}$ which contains the generalized eigenvectors of the pair $\big\{\breve{\textbf{H}}_{k,l,k,l}^H\breve{\textbf{H}}_{k,l,k,l},~\tilde{\textbf{H}}_{k,l}^H\tilde{\textbf{H}}_{k,l}+\gamma\textbf{I}_{M_\T^{\RF}}\big\}$. $\textbf{W}_{\BB_{k,l}}$ is designed as a matched filter at the receiver~\cite{Sadek07}:
\begin{equation}\label{eq:WBB_SLNR}
\textbf{W}_{\BB_{k,l}}=\frac{\breve{\textbf{H}}_{k,l,k,l}\textbf{F}_{\BB_{k,l}}}{||\breve{\textbf{H}}_{k,l,k,l}\textbf{F}_{\BB_{k,l}}||_F}
\end{equation}
\vspace{-4 mm}
\subsection{Generalized Maximum-Ratio Precoding}
\textcolor{black}{The fourth HBF strategy is generalized maximum-ratio (GMR) transmission that belongs to coordinated scheduling/beamforming in CoMP, and} has the same RF precoding, RF combining, and baseband combining procedures as the SLNR-based approach. In the GMR-based method, the effective channel for user $k$ in cell $l$ after RF precoding and RF combining is denoted as the $N_\R^{\RF}\times M_\T^{\RF}$ matrix $\check{\textbf{H}}_{m,i,k,l}$ defined as:
\begin{equation}\label{eq:HcZF}
\check{\textbf{H}}_{m,i,k,l}=\frac{1}{\sqrt{\PL_{m,i,l}}}\textbf{W}_{\RF_{m,i}}^H\textbf{H}_{m,i,l}\textbf{F}_{\RF_{k,l}}
\end{equation} 

\noindent and the $KLN_\R^{\RF}\times M_\T^{\RF}$ concatenated effective channel matrix is:
\begin{equation}\label{eq:HtZF}
\tilde{\textbf{H}}_{k,l}=[\check{\textbf{H}}_{1,1,k,l}^T, ..., \check{\textbf{H}}_{k,l,k,l}^T,..., \check{\textbf{H}}_{K,L,k,l}^T]^T
\end{equation}

\noindent If $N_\R^{\RF}=N_\S$, the baseband precoding matrix can be set as the $N_\S(K(l-1)+k-1)+1$th to the $N_\S(K(l-1)+k)$th columns of $\overline{\textbf{F}_\BB}$ yielded by the GMR transmission matrix:
\begin{equation}
\overline{\textbf{F}_\BB} = \tilde{\textbf{H}}_{k,l}^H
\end{equation}

\noindent Or equivalently
\begin{equation}\label{eq:FBBGMR}
\textbf{F}_{\BB_{k,l}} = \check{\textbf{H}}_{k,l,k,l}^H
\end{equation}

\noindent Eq.~\eqref{eq:FBBGMR} shows that GMR essentially requires no coordination among TPs. However, it should be noted that GMR only works for the situation where $N_\R^{\RF}=N_\S$, and will not work otherwise due to matrix dimension mismatch. All the other proposed schemes work for any situations where $N_\R^{\RF}\geq N_\S$. \textcolor{black}{In practice, the dimension issue is easily accounted for by turning off the unnecessary RF chains.}
\vspace{-3 mm}
 \textcolor{black}{\subsection{Feasibility of Zero-Forcing Precoding}
 Another popular multi-user precoding method besides maximum ratio (MR) is zero-forcing (ZF)~\cite{Zakhour12}, thus it is reasonable to consider whether ZF precoding is feasible in the system setup herein. Analogous to GMR introduced in the previous subsection, let us assume the RF precoding, RF combining, and baseband combining schemes are the same as those in the GMR-based HBF method, and that $N_\S=N_\R^{\RF}$, then the baseband precoding matrix for user $k$ in cell $l$ $\textbf{F}_{\BB_{k,l}}$ is composed of the $N_\S(K(l-1)+k-1)+1$th to the $N_\S(K(l-1)+k)$th columns of $\overline{\textbf{F}_\BB}$ given by the generalized ZF matrix:
\begin{equation}
\overline{\textbf{F}_\BB} = \tilde{\textbf{H}}_{k,l}^H(\tilde{\textbf{H}}_{k,l}\tilde{\textbf{H}}_{k,l}^H)^{-1}
\end{equation}}
\vspace{-1 mm}
\noindent\textcolor{black}{where $\tilde{\textbf{H}}_{k,l}$ is given by~\eqref{eq:HtZF} with the dimension $KLN_\R^{\RF}\times M_\T^{\RF}$, hence $\tilde{\textbf{H}}_{k,l}\tilde{\textbf{H}}_{k,l}^H$ has the dimension $KLN_\R^{\RF}\times KLN_\R^{\RF}$ with a rank of $M_\T^{\RF}$ which is smaller than $KLN_\R^{\RF}$. Therefore, $\tilde{\textbf{H}}_{k,l}\tilde{\textbf{H}}_{k,l}^H$ is rank deficient thus not invertible, hence ZF precoding is not feasible for the proposed multi-cell system due to dimension constraints. Alternatively, the rank deficiency problem will not exist if ZF is done at the receiver side, which, however, requires that each user has the CSI of all TPs to all users, and this is too much overhead for the user hence not feasible, either. While regularized ZF (RZF) can be used to avoid the rank deficiency issue in ZF, the optimal regularization parameter remains to be solved for multi-cell multi-stream scenarios, which is outside the scope of this paper. Further, the performance of RZF approximates MR for low SNRs and ZF for high SNRs~\cite{Sun_PhDThesis}, thus MR and ZF are sufficiently instructive.}
\vspace{-2 mm}
\section{Channel Model Parameter Setting}
\vspace{-1.27 mm}
Two types of channel models that can be regarded as promising candidates for 5G wireless \textcolor{black}{system simulation} are the 3GPP TR 38.901 Release 14 channel model~\cite{3GPP_38.901} and NYUSIM channel model~\cite{Sun17_NYUSIM,Rap17_Tech}. The former is inherited from sub-6 GHz communication system models with modifications to accommodate the spectrum above 6 GHz up to 100 GHz\textcolor{black}{~\cite{5GCM}}. The NYUSIM model is \textcolor{black}{also} developed based on extensive real-world propagation measurements at multiple mmWave frequency bands and is able to faithfully reproduce the channel impulse responses obtained from over 1 Terabytes of measured data~\cite{Rap15:TCOM,Samimi15:MTT,Sun17_NYUSIM}. Both 3GPP and NYUSIM models include basic channel model components such as line-of-sight probability model, large-scale path loss model, large-scale parameters, small-scale parameters, etc. However, the approaches and parameter values used in each modeling step can be significantly different. Both 3GPP TR 38.901 Release 14~\cite{3GPP_38.901} and NYUSIM~\cite{Sun17_NYUSIM} models will be used to investigate the impact of different channel models on the multi-cell HBF performance, where the frequency domain representation, (i.e. complex gains for each OFDM channel across the spectrum)\textcolor{black}{~\cite{Rap15}} is applied in space across the antenna manifold at a single epoch for analysis. Channel model parameter settings utilized in the simulations are given in Table~\ref{tbl:SimSet}. 
\vspace{-2 mm}
\section{Simulation Results and Analysis}
\begin{figure}
	\centering
	\includegraphics[width=3.0in]{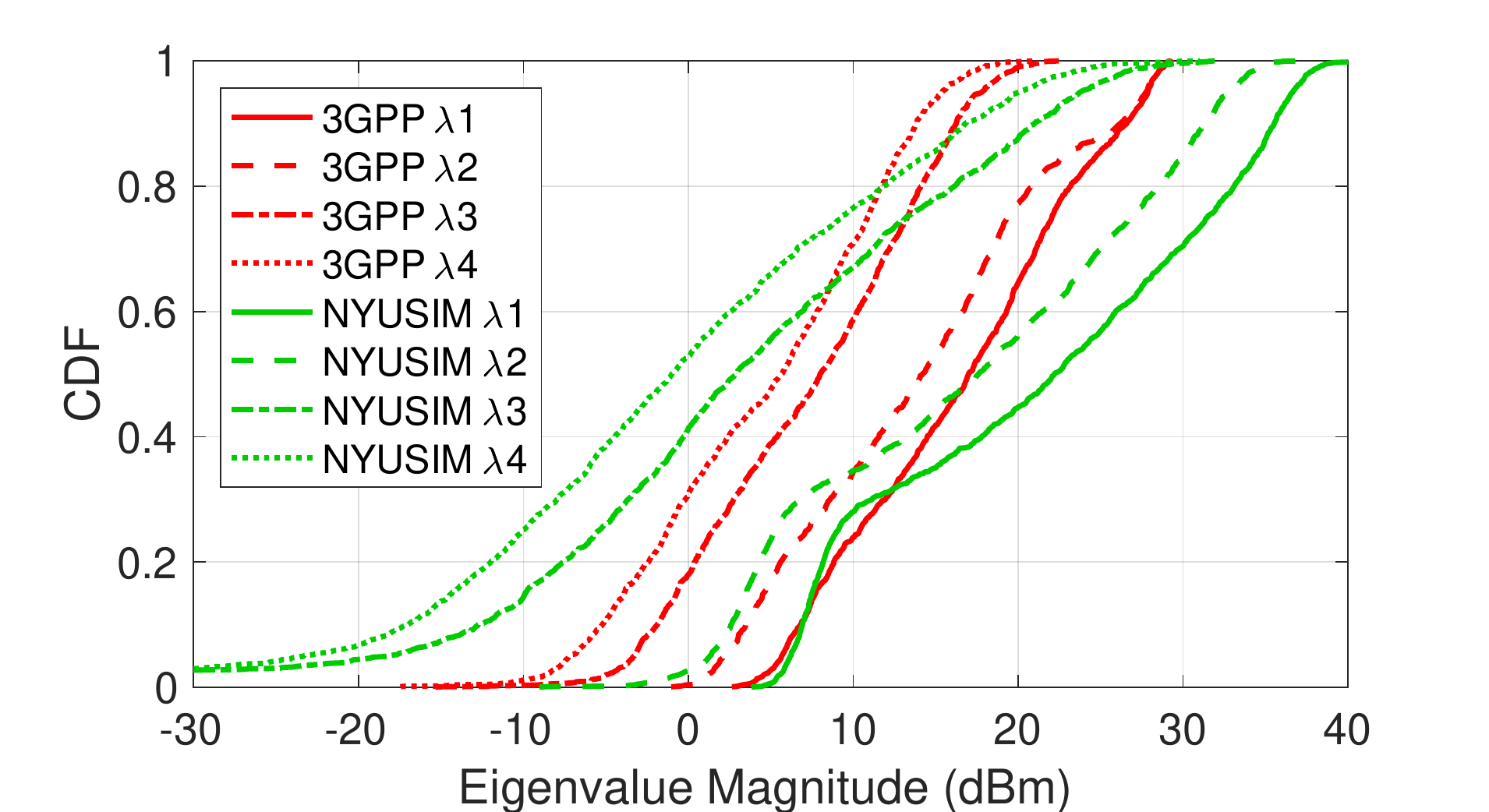}
	\vspace{-3 mm}
	\caption{CDFs of the largest four \textcolor{black}{eigenvalues of $\textbf{H}\textbf{H}^H$} in 3GPP and NYUSIM channel models for each individual user in a three-cell three-user MIMO\textcolor{black}{-OFDM} system in the UMi scenario. The transmit and receive antenna arrays are URAs composed by 256 and 8 cross-polarized elements, respectively. The carrier frequency is 28 GHz with an RF bandwidth of 100 MHz \textcolor{black}{with narrowband frequency-flat-fading sub-carriers}. Each TP antenna element has a radiation pattern as specified in Table 7.3-1 of~\cite{3GPP_38.901} with a maximum gain of 8 dBi, and each UE antenna element possesses an omnidirectional pattern.}
	\label{fig:EV}	
	\vspace{-4.5 mm}
\end{figure} 
\begin{figure*}
	\centering
	\subfloat[50 m cell radius, 12 users per cell, two streams per user]{%
		\includegraphics[width=3in]{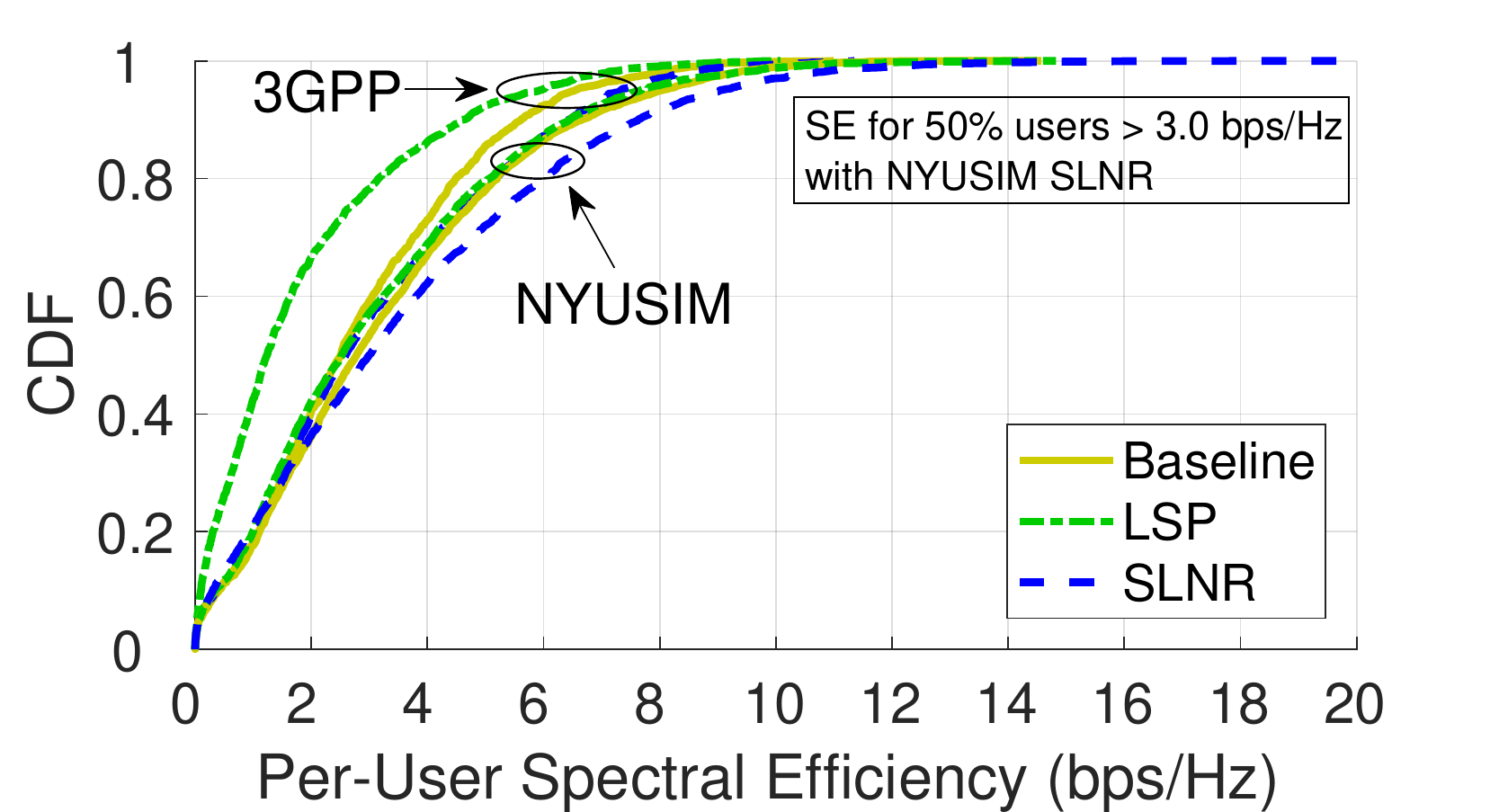}}	
	\subfloat[50 m cell radius, three users per cell, two streams per user]{%
		\includegraphics[width=3in]{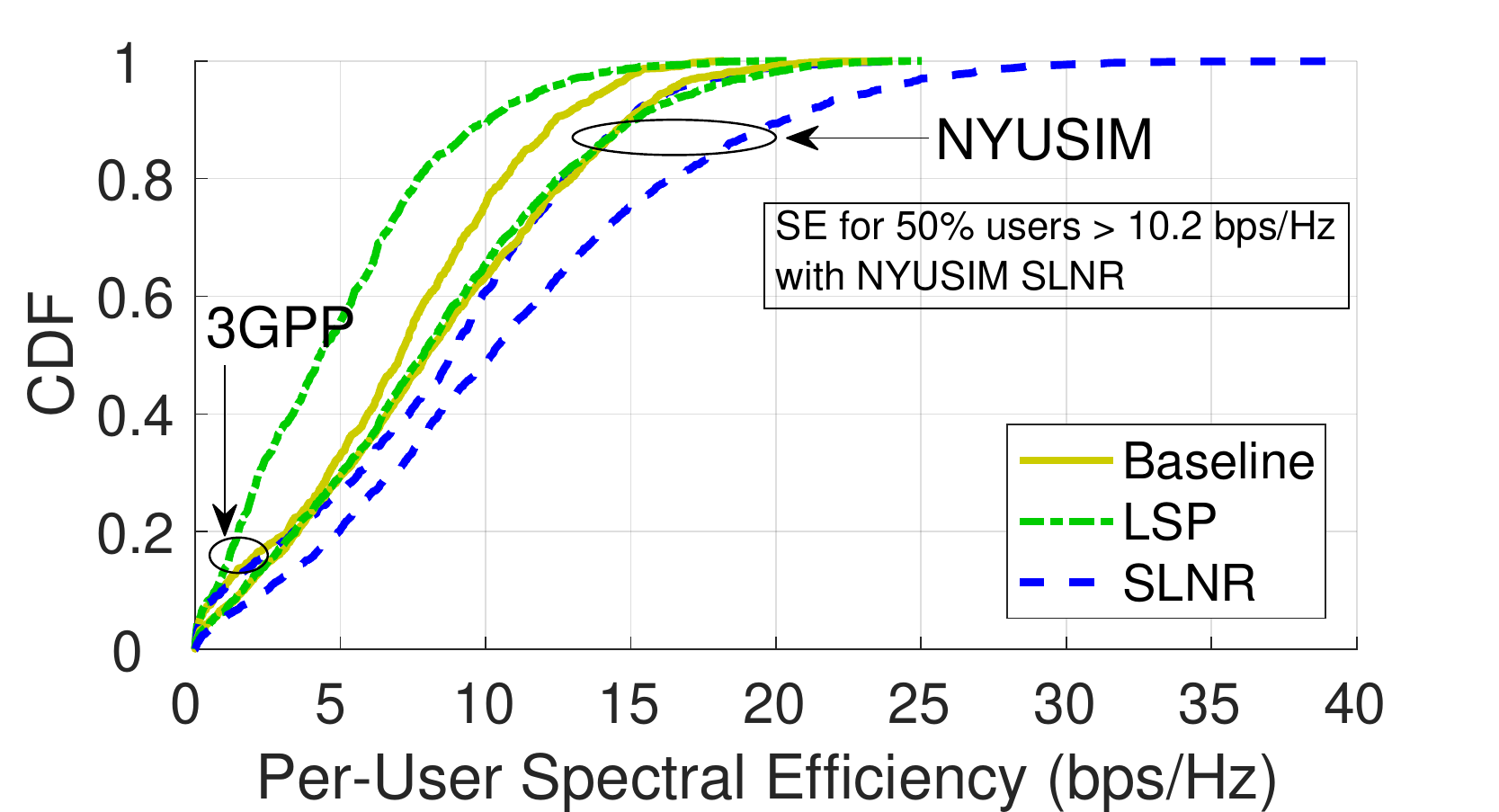}}	
	
	\subfloat[200 m cell radius, 12 users per cell, two streams per user]{%
		\includegraphics[width=3in]{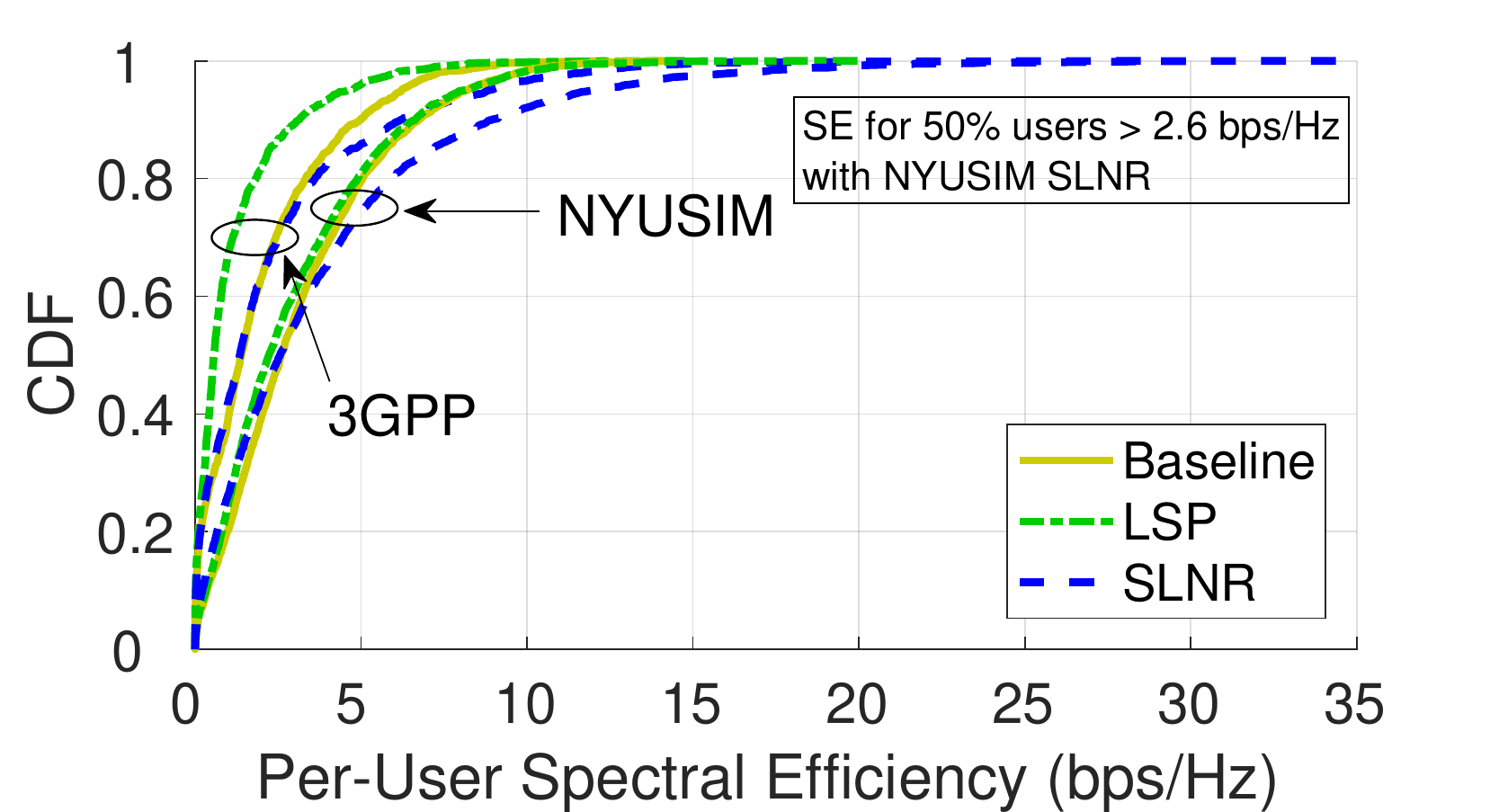}}
	\subfloat[200 m cell radius, three users per cell, two streams per user]{%
		\includegraphics[width=3in]{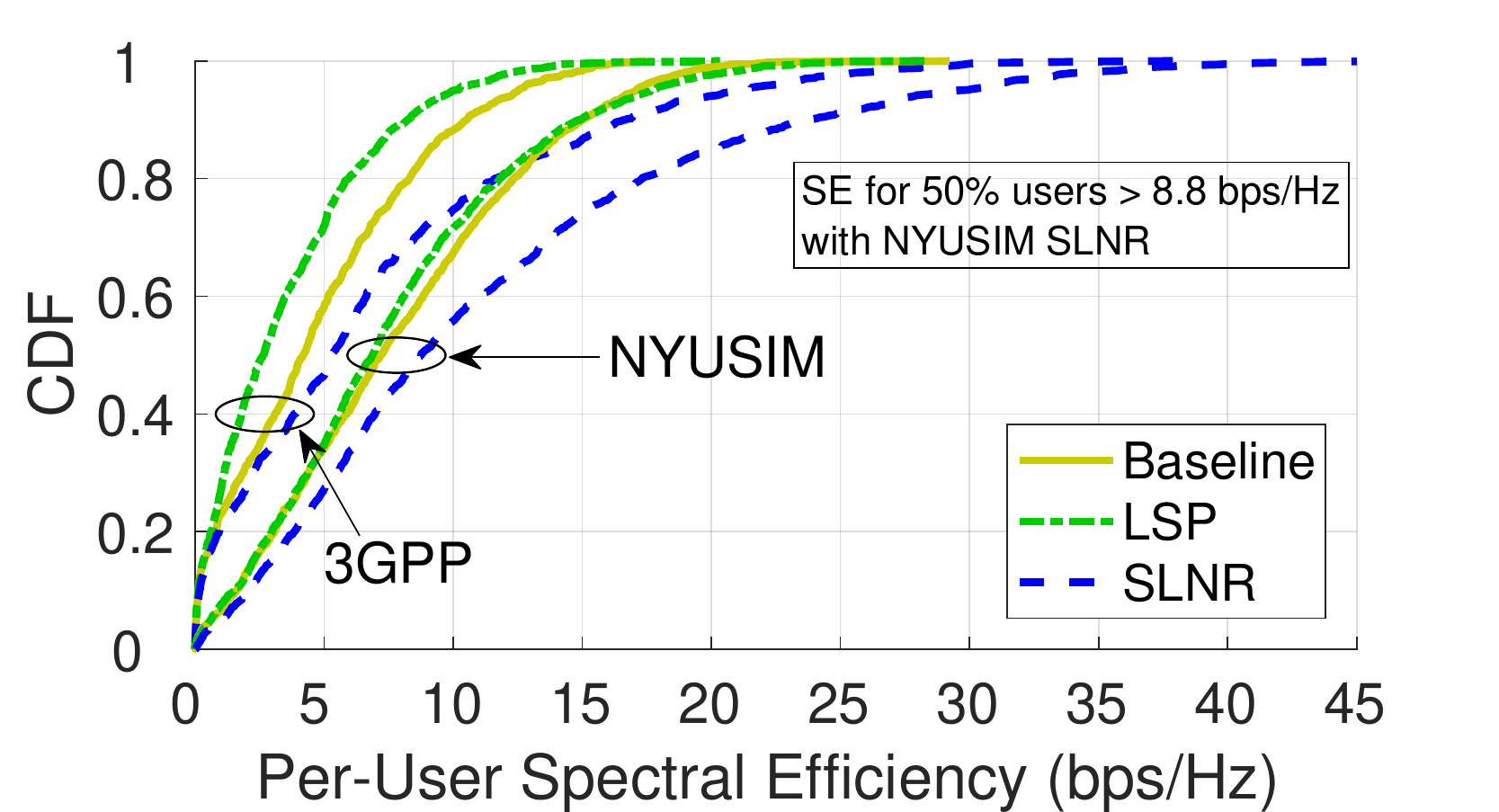}}
	\vspace{-1.5 mm}
	\caption{CDFs of the spectral efficiency per user with (a) a 50 m cell radius and 12 users per cell, (b) a 50 m cell radius and three users per cell, (c) a 200 m cell radius and 12 users per cell, and (d) a 200 m cell radius and three users per cell, in the three-cell MIMO system using the HBF approaches proposed in this paper for 3GPP~\cite{3GPP_38.901} and NYUSIM~\cite{Sun17_NYUSIM} channel models. There is one TP per cell, four RF chains \textcolor{black}{and two streams} per user, and 48 and 12 TP RF chains for 12 and three users per cell, respectively.}
	\label{fig:MCSE}	
	\vspace{-3 mm}
\end{figure*}
\begin{figure}
	\centering
	\includegraphics[width=2.8in]{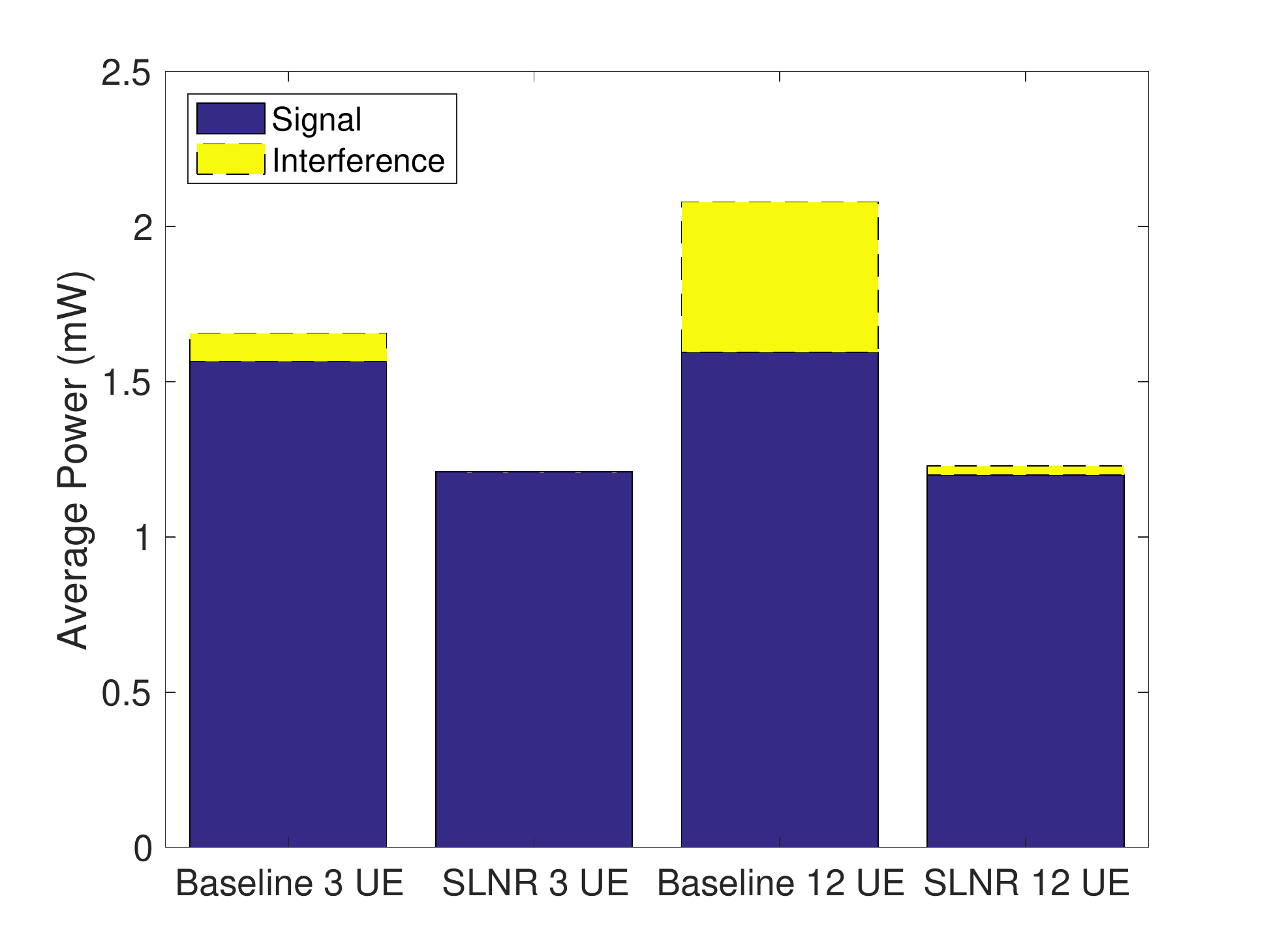}
	\vspace{-6 mm}
	\caption{Average signal power and interference power generated from the NYUSIM channel model for the three-cell system with a cell radius of 50 m, where the average is taken over users. There are two streams and four RF chains per user, and 48 and 12 TP RF chains for 12 and three users per cell, respectively.}
	\label{fig:PowerBar}	
	\vspace{-5 mm}
\end{figure}
\begin{figure}
	\centering
	\subfloat[Two streams per user, 50 m cell radius, three users per cell]{%
		\includegraphics[width=3.7in]{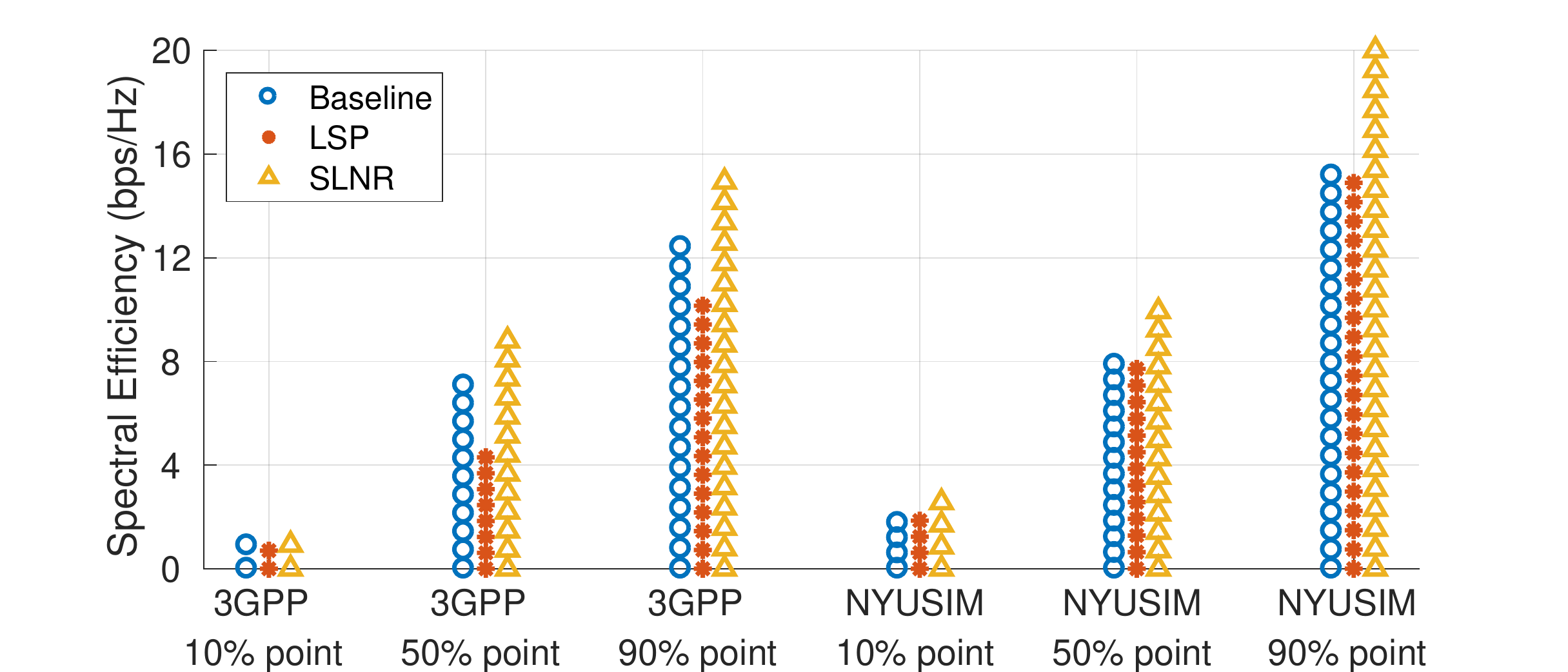}}
\vspace{-2 mm}	
	\subfloat[Four streams per user, 50 m cell radius, three users per cell]{%
		\includegraphics[width=3.7in]{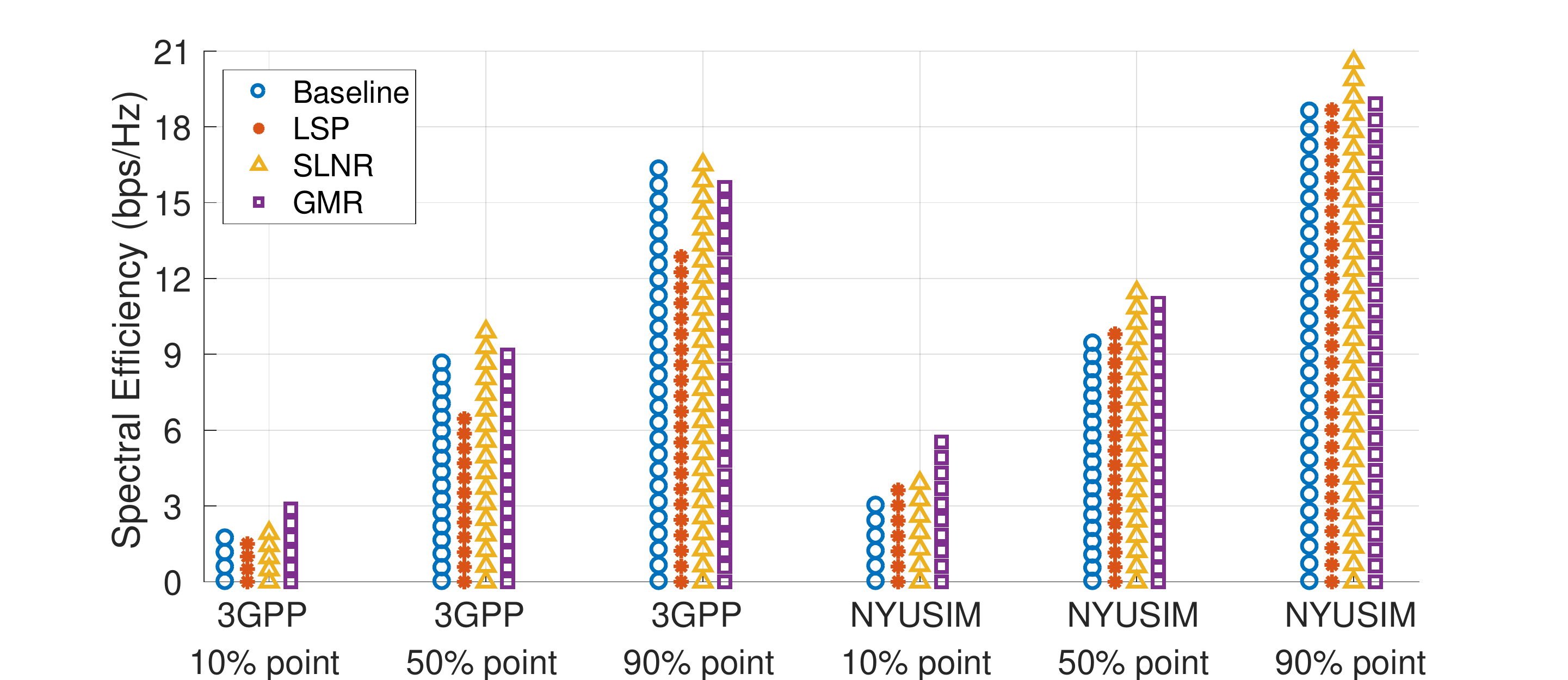}}
	\vspace{-1 mm}
	\caption{CDFs of the per-user spectral efficiency of the three-cell multi-user MIMO system using the HBF approaches proposed in this paper for 3GPP~\cite{3GPP_38.901} and NYUSIM~\cite{Sun17_NYUSIM} channel models for the cases of (a) two streams, and (b) four streams per user. The users in each cell are distributed uniformly and randomly with T-R separation distances ranging from 10 m to 50 m.}
	\label{fig:MCSEBar}	
	\vspace{-2 mm}
\end{figure}
\vspace{-0 mm}
\subsection{\textcolor{black}{Eigenvalues of $\textbf{H}\textbf{H}^H$}}
\vspace{-1 mm}
\textcolor{black}{Eigenvalues of $\textbf{H}\textbf{H}^H$ are} a measure of the power contained in eigenchannels for spatial multiplexing in a \textcolor{black}{MIMO-OFDM} system. We generate the downlink $N_\R\times N_\T$ MIMO channel matrix \textbf{H} using both 3GPP~\cite{3GPP_38.901} and NYUSIM~\cite{Samimi15:MTT,Sun17_NYUSIM} channel models, for a system operating at 28 GHz with 100 MHz RF bandwidth \textcolor{black}{and narrowband frequency-flat fading sub-carriers}, and 256 antennas \textcolor{black}{in the TP URA} and eight antennas \textcolor{black}{in the UE URA}. \textcolor{black}{Although the channel coefficients in $\textbf{H}$ over the 100 MHz usually vary with carrier frequency, the mean values (statistics) of the eigenvalues of $\textbf{H}\textbf{H}^H$, where the superscript $H$ denotes conjugate transpose, are generally frequency-independent over the 100 MHz bandwidth. In other words, the narrowband flat fading will be identical in statistics at any sub-carrier in the 100 MHz RF channel bandwidth, so for simplicity, we use the channel impulse response from the 3GPP channel model and the NYUSIM channel model, respectively, and apply the resulting narrowband complex channel gain/channel state at the center frequency sub-carrier of 28.000 GHz.} Fig.~\ref{fig:EV} depicts the cumulative distribution functions (CDFs) of the largest four eigenvalues of \textbf{H}\textbf{H}$^H$ for both 3GPP~\cite{3GPP_38.901} and NYUSIM~\cite{Samimi15:MTT,Sun17_NYUSIM} models for each individual user in a three-cell three-user \textcolor{black}{MIMO-OFDM} system in the urban microcell (UMi) scenario. Fig.~\ref{fig:EV} shows that the highest two eigenvalues \textcolor{black}{of $\textbf{H}\textbf{H}^H$} in NYUSIM are larger than those in 3GPP in most cases, while the third and fourth eigenvalues are smaller most of the time. This indicates that NYUSIM yields only a few but strong dominant eigenmodes, whereas the 3GPP model generates more eigenmodes with weaker powers. The number of dominant eigenchannels (i.e., the channel rank) in NYUSIM is statistical and can vary over the range of 1 to 5, where 5 is the maximum number of spatial lobes~\cite{Samimi15:MTT}, with an average and typical value of 2 over numerous simulations.
\vspace{-3 mm}
\subsection{Spectral Efficiency}
Using the multi-cell multi-user MIMO (MU-MIMO) HBF procedures proposed above and the three-cell layout demonstrated in Section~\ref{sec:MCSL}, and the simulation settings shown in Table~\ref{tbl:SimSet}, spectral efficiency is studied using both the 3GPP and NYUSIM channel models via MATLAB simulations. For each channel model, 400 random channel realizations were carried out where 27 channel matrices were generated in each channel realization for the three-user-per-cell case (hence resulting in 10800 channel matrices in total), which represent the channel matrices between each TP and each UE in the three cells; while 100 random channel realizations were carried out where 108 channel matrices were generated in each channel realization for the 12-user-per-cell case (hence resulting in 10800 channel matrices in total). In each channel realization, UE locations in each cell are randomly and uniformly generated with T-R separation distances ranging from 10 m to the cell radius. The cell radius is set to 50 m and 200 m, respectively, where the 200 m radius is obtained by assuming that 95\% of the area in each cell has an SNR larger than or equal to 5 dB, and the upper bound of the T-R separation distance is calculated based on this assumption and is rounded to 200 m for both models for fair comparison~\cite{3GPP_38.901,Sun17_NYUSIM}, while the 50 m radius is chosen for comparison purposes.

The CDFs of per-user spectral efficiency in the three-cell MU-MIMO system using both 3GPP~\cite{3GPP_38.901} and NYUSIM~\cite{Sun17_NYUSIM} models are illustrated in Fig.~\ref{fig:MCSE} for different cell radii and \textcolor{black}{numbers of users} with two steams per user. Fig.~\ref{fig:MCSE} shows that for both 3GPP and NYUSIM models, the SLNR-based HBF outperforms all the other HBF schemes, revealing its effectiveness in suppressing both intra-cell and inter-cell interference and noise. Another distinguishing feature is that LSP does not outperform the baseline case for the 3GPP model, which is probably due to the fact that LSP spends part of the transmit power on suppressing leakage, thus leaving less power for signal transmission compared to the baseline case. \textcolor{black}{In contrast, LSP works much better using the realistic NYUSIM channel model (up to 150\% improvement than using the 3GPP channel model for 50\% of users),} since the NYUSIM channel has a stronger dominant eigenchannel than 3GPP (see Fig.~\ref{fig:EV}), thus LSP appears \textcolor{black}{to be much} more effective \textcolor{black}{when using the NYUSIM channel model, since} the dominant leakage is stronger. Furthermore, \textcolor{black}{using NYUSIM leads to} higher spectral efficiency as compared to the 3GPP model, likely due to the stronger two dominant eigenchannels per user yielded by NYUSIM channel matrices. 

When comparing Figs.~\ref{fig:MCSE}(a) and~\ref{fig:MCSE}(b), or Figs.~\ref{fig:MCSE}(c) and~\ref{fig:MCSE}(d), it is noticeable that for the same cell radius, the spectral efficiency gap between the SLNR approach and the baseline decreases as the \textcolor{black}{number of users} increases. This phenomenon can be explained by Fig.~\ref{fig:PowerBar} which depicts the average signal power and interference power (averaged over users) for different \textcolor{black}{numbers of users} using the SLNR method and the baseline for the 50 m cell radius as an example. Fig.~\ref{fig:PowerBar} shows that for either the SLNR approach or the baseline, when the \textcolor{black}{number of users} increases from three to 12, the average signal power remains almost the same, while the average interference power increases, and the ratio of the interference power in the baseline to that in the SLNR scheme is smaller in the 12-user case than in the three-user case (about 16 versus 140), since the interference power in the SLNR method approaches zero for the three-user case. Therefore, the corresponding SINR gap and hence the spectral efficiency gap is smaller in the 12-user case. 

Moreover, it is observable by comparing Figs.~\ref{fig:MCSE}(a) and~\ref{fig:MCSE}(c), or Figs.~\ref{fig:MCSE}(b) and~\ref{fig:MCSE}(d), that for the majority (about 70\%-90\%) of the users, the spectral efficiency for the 200 m cell radius is lower than the 50 m cell radius for any of the proposed HBF schemes with the same \textcolor{black}{number of users} per cell and the same transmit power per user, except for the peak spectral efficiency. This indicates that the effect of interference does not dictate the spectral efficiency, \textcolor{black}{but rather coverage/SNR matters most,} since the 200 m cell radius corresponds to weaker interference but has lower spectral efficiency in most cases.

Next, we consider the case where each TP communicates with each of its home-cell users via four data streams, along with the two-stream-per-user case. As $N_\S=N_\R^{\RF}$ in the four-stream-per-user case, GMR is \textcolor{black}{tractable} hence is considered herein. Fig.~\ref{fig:MCSEBar} depicts the 10\%, 50\%, and 90\% CDF points of spectral efficiency for both 3GPP and NYUSIM models for two-stream and four-stream cases with a cell radius of 50 m and three users per cell. As unveiled by Fig.~\ref{fig:MCSEBar}, SLNR yields the highest spectral efficiency except for the 10\% CDF point in Fig.~\ref{fig:MCSEBar}(b), where GMR outperforms all the other HBF schemes since GMR intrinsically maximizes the received signal power hence is more efficient when the SNR is low. Interestingly, the eigenmode beamforming scheme in the baseline case exhibits better performance as the number of streams increases, especially for the 3GPP channel model, likely due to its capability to focus all the transmit power onto strongest eigenmodes, and that the third and fourth eigenmodes in the 3GPP model are mostly stronger than those in NYUSIM (see Fig.~\ref{fig:EV}). Figs.~\ref{fig:MCSE} and~\ref{fig:MCSEBar} indicate that \textcolor{black}{CoMP} (e.g., SLNR) generally provides higher spectral efficiency than the \textcolor{black}{non-CoMP} case \textcolor{black}{(e.g., up to 67\% more spectral efficiency for the weakest 5\% of users using SLNR-based CoMP)}, thus is worth using in mmWave multi-cell networks.
\vspace{-3 mm}
\section{Conclusions} 
\vspace{-1 mm}
In this paper, we considered multi-cell multi-user \textit{multi-stream} communication in mmWave homogeneous networks, and proposed and compared four HBF approaches based on the assumption that TPs in different cells have full CSI and can exchange the CSI among each other, such that the TPs can take into account both intra-cell and inter-cell interference when designing precoding matrices. Numerical results show that \textcolor{black}{SLNR-based CoMP provides highest spectral efficiency in most cases (e.g., up to 67\% higher spectral efficiency for the weakest 5\% of users as compared to the non-CoMP case),} thus is worth using in mmWave multi-cell networks. \textcolor{black}{LSP shows minimal improvement over the baseline, and ZF is not feasible due to rank deficiency of the product of effective channel matrices after RF precoding and combining.} Moreover, the behaviors of the four proposed multi-stream HBF approaches are affected by the interference and SNR level, which are themselves influenced by the cell radius, the number of users per cell, and the number of streams per user. Specifically, a \textcolor{black}{relatively small cell radius (e.g., 50 m) and a small number of users (e.g., three)} per cell usually give rise to high per-user spectral efficiency given a constant transmit power for each user.
\vspace{-0 mm}
\bibliographystyle{IEEEtran}
\vspace{-5.8 mm}
\bibliography{Multi-Cell}

\end{document}